\documentclass[12pt]{article}
\usepackage{epsfig}

\def\beq{\begin{equation}}
\def\eeq#1{\label{#1}\end{equation}}
\def\eeqn{\end{equation}}

\def\beqa{\begin{eqnarray}}
\def\eeqa#1{\label{#1}\end{eqnarray}}
\def\eeqan{\end{eqnarray}}

\let\bar=\overbar

\def\Dslash{\not{\hbox{\kern-4pt $D$}}}
\def\dslash{\not{\hbox{\kern-2pt $\del$}}}

\def\msb{{\bar{\ssstyle M \kern -1pt S}}}

\def\Title#1{\begin{center} {\Large {\bf #1} } \end{center}}

\begin{document}

\Title{The Construction and Commissioning of the 
Belle~II iTOP Counter}

\bigskip\bigskip


\begin{raggedright}  

{\it Boqun Wang\index{Wang, B.}, on behalf of the Belle~II iTOP
group\\ Department of Physics\\ University of Cincinnati\\ Cincinnati,
Ohio 45221, USA}

\bigskip

Talk presented at the APS Division of Particles and Fields Meeting
(DPF 2017), July 31-August 4, 2017, Fermilab. C170731

\bigskip\bigskip
\end{raggedright}

\begin{abstract}
The barrel-region particle identification detector is crucial for
extending the physics reach of the Belle~II experiment operating at
the SuperKEKB accelerator. For this purpose, an imaging-Time-of-
Propagation (iTOP) counter was developed, which is a new type of ring-
imaging Cherenkov detector. The iTOP consists of 16 separate modules
arranged azimuthally around the beam line.  Each module consists of
optical components fabricated from quartz (one mirror, one prism,  and
two bars), an array of micro-channel-plate photo-multiplier tubes
(MCP-PMTs), and front-end electronics. The waveforms read out are
processed by firmware, and the resulting pulse-heights and hit times
are sent to the Belle~II data acquisition system. The detector
construction was completed and the detector installed by the summer of
2016, and since then the detector has undergone commissioning. This
talk describes the construction and commissioning of the Belle~II iTOP
counter.
\end{abstract}

\section{Introduction}

The B factory is an $e^+e^-$ collider running at the $\Upsilon(4S)$
resonance energy to produce B meson pairs. The major B factories are
Belle running at KEKB in Japan and BaBar running at PEP-II in US.
These two facilities have collected $\sim 1.5$ $ab^{-1}$ in total of
$e^+e^-$ collision data. With this data set, they've reached physics
achievements in areas like the CKM angle measurement, $|V_{cb}|$ and
$|V_{ub}|$ measurement, semileptonic and leptonic B decays, rare B
decays, $\tau$ physics, $D^0$ mixing and CP violation, $B_s$ physics
at the $\Upsilon(5S)$, two-photon physics and new resonances.

The Belle~II~\cite{belle2} detector at the SuperKEKB~\cite{superkekb}
accelerator is an upgrade of the Belle detector at the KEKB
accelerator for searching for New Physics (NP), which is physics
beyond the Standard Model (SM). The design luminosity of the upgraded
accelerator is $8 \times 10^{35} cm^{-2} s^{-1}$, which is $\sim$ 40
times larger than that of the KEKB collider. The nano-beam
technology~\cite {nano-beam} is utilized to significantly squeeze the
sizes of the beam bunches to achieve such high luminosity. The planned
integrated luminosity taken by the Belle~II detector is $\sim 50$
$ab^{-1}$, which is around 50 times larger than that of Belle.

The iTOP (imaging-Time-Of-Propagation) counter~\cite{itop-inami,  itop-matsuoka, itop-inami2, itop-boqun}, which is the particle
identification detector in the barrel region, is a newly designed
sub-detector for the Belle~II detector. It consists of a 2.7 m long
quartz optics for the radiation and propagation of the Cherenkov
light, an array of micro-channel- plate photo-multiplier tubes        (MCP-PMT)~\cite{mcppmt} for photon detection, and wave-sampling front-end
readout electronics~\cite{itop-irsx, itop-irsx1}. This article
describes the construction and commissioning of the iTOP counter.

\section{Detector Design}

\begin{figure}[tb]
	\centering
	\includegraphics[width=0.6\textwidth]{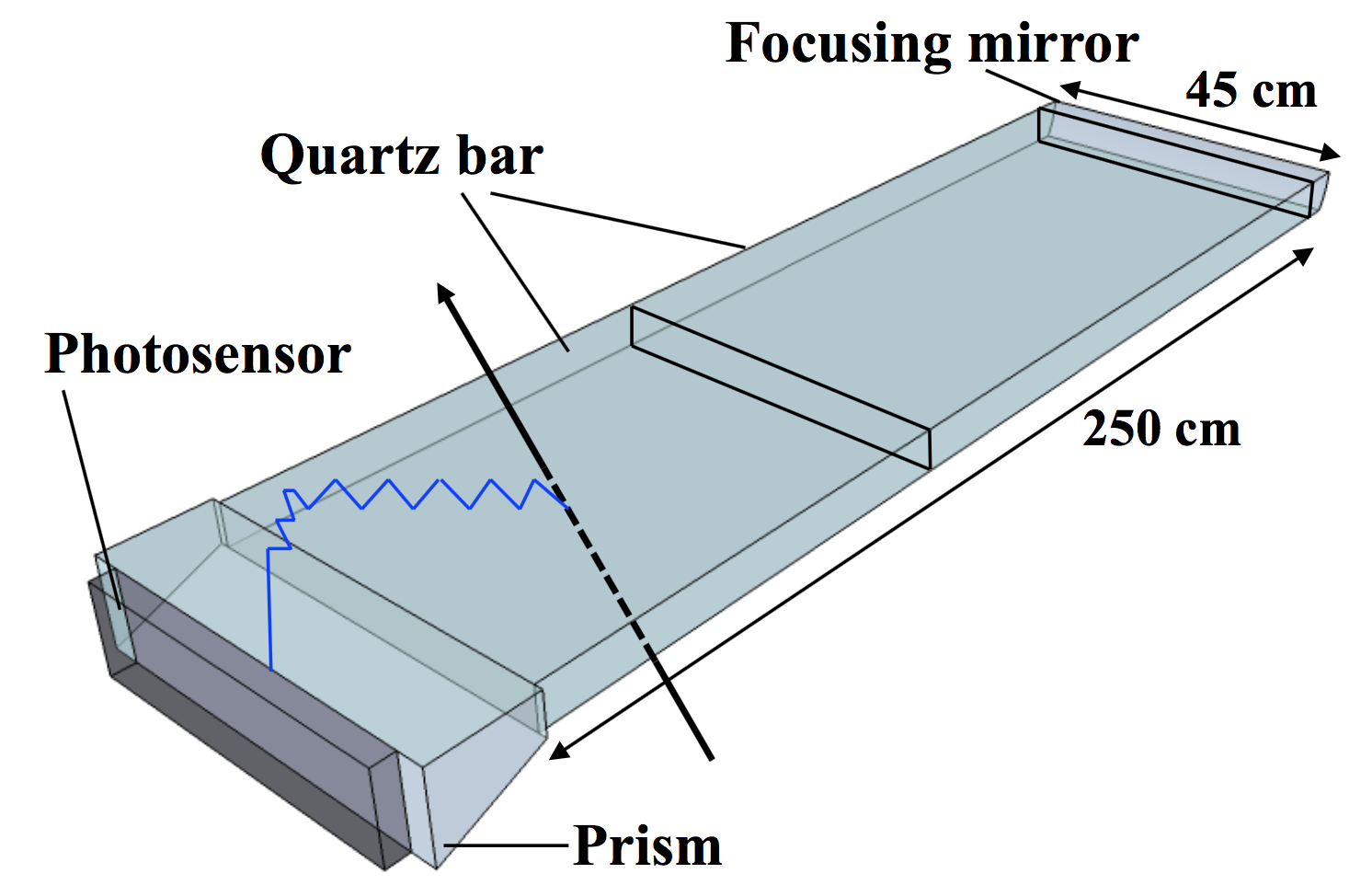}
	\caption{Optical overview of the iTOP counter.}
	\label{fig:itop}
\end{figure}

As shown in Figure~\ref{fig:itop}, one iTOP module consists of two
1250 mm long, 450 mm wide and 20 mm thick bars. A reflective mirror
with spherical surface is mounted at one end of the bars, and an
expansion block called prism is mounted at the other end. All optics
components are made of Corning 7980 synthetic fused silica, which has
a high purity and no striae inside. There are 16 modules in total for
the Belle~II detector.

\begin{figure}[tb]
	\centering
	\includegraphics[width=0.6\textwidth]{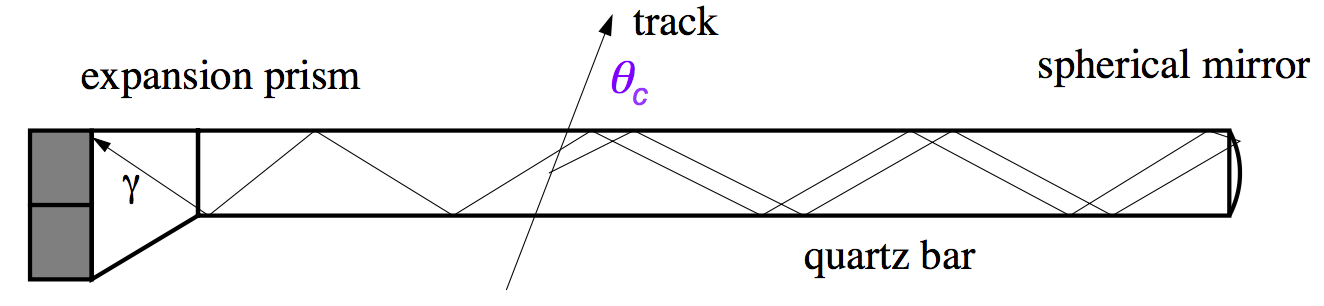}
	\caption{The working principle of the iTOP counter.}
	\label{fig:itop-work}
\end{figure}

The diagram of the working principle of the iTOP counter is shown in
Figure~\ref{fig:itop-work}. The Cherenkov photons are emitted when a
charged track goes through the quartz radiator, and the Cherenkov
angle depends on the velocity of the charged track. With the momentum
of the charged track given by the central drift chamber (CDC), the
angle depends on the mass of the charged particle. The photons are
collected by the MCP-PMTs after being reflected by the bar surfaces
and the mirror. The resolution of the photon sensors and the front end
electronics are required to be better than 50 ps, which is required to
distinguish the time of propagation difference between Cherenkov
photons from $\pi^{\pm}$ and $K^{\pm}$ tracks.

\section{Module Construction}

The construction process of the iTOP modules was started at the end of
2014 and all 17 modules, including one spare, were finished by April
2016. After testing with laser and cosmic rays, these modules were
installed in the Belle~II detector by May 2016.

\subsection{QA of Quartz Optics}

The optics of the iTOP counter needs to have very high optical quality
to achieve the high K/$\pi$ separation capability. The Cherenkov
photons will be reflect $\sim$ 50 -- 100 times inside the quartz
radiator, so the surfaces of the quartz bars need to be highly
polished. The requirement for surface roughness is $<$ 5 \AA ~r.m.s.,
and for flatness the requirement is $<$ 6.3 $\mu m$. For all 34 bars
needed, 30 were produced by Zygo Corporation (USA) and 4 were produced
by Okamoto Optics Works (Japan).

\begin{figure}
  \centering
  \includegraphics[width=0.6\textwidth]{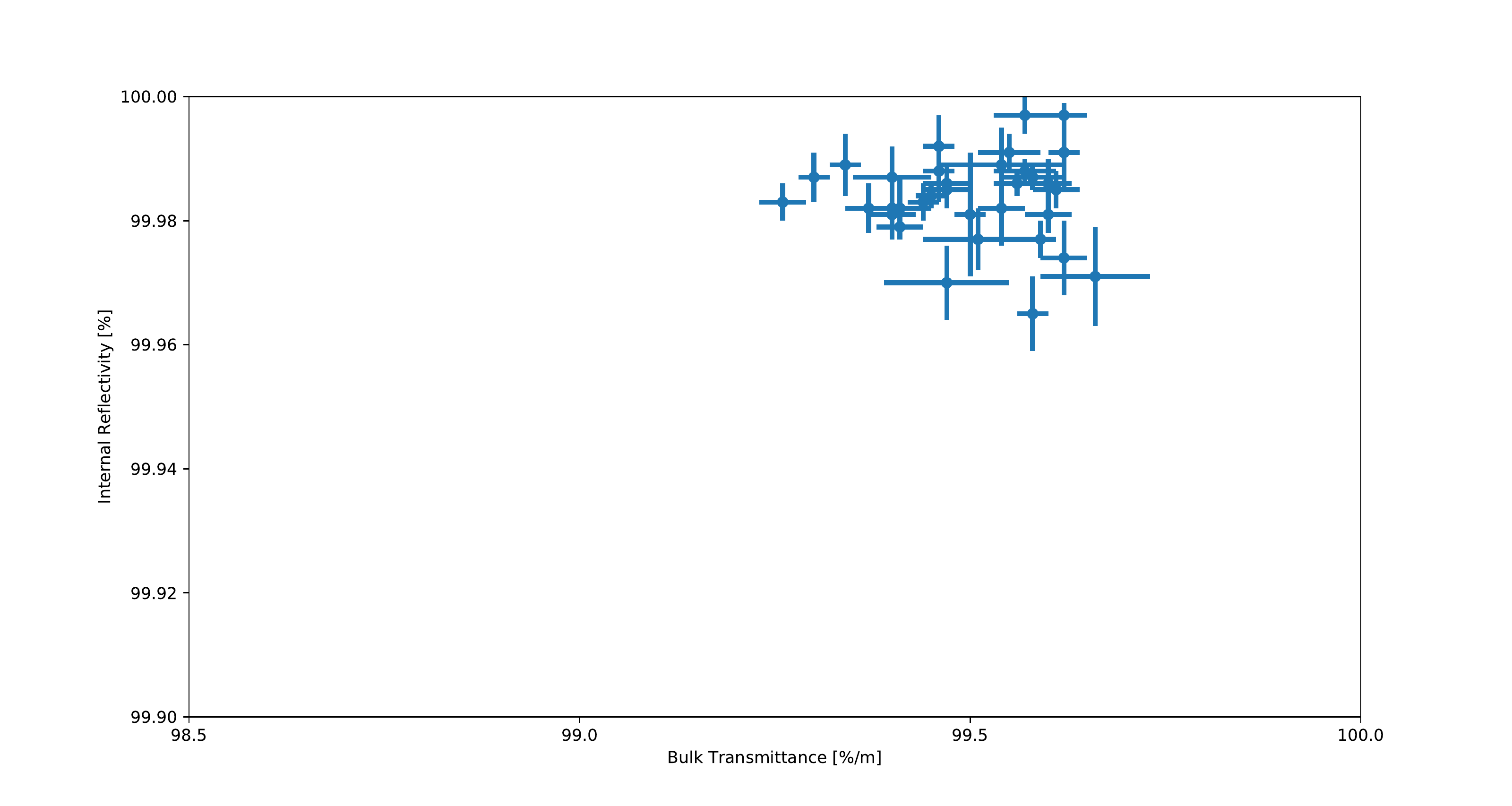}
  \caption{Summary of quartz bar QA results.}
  \label{fig:bar-qa}
\end{figure}

The quartz bars were procured by the vendors and delivered directly to
KEK. After receiving these bars, they were mounted on the
measurement stage for the QA (Quality Assurance) tests. Laser beam was
injected to one surface of the bar and went out from the other
surface. By measuring the beam intensity before and after it went
through the quartz bar, the bulk transmission can be measured. For the
internal reflectivity measurement, the methods were similar except the
laser beam was injected with a certain angle. The requirements for
bulk transmittance and internal reflectivity were $>$ 98.5 \%/m and
$>$ 99.9 \% per reflection, respectively. As shown in Figure~\ref{fig:bar-qa}, all received quartz bars fulfilled the requirements.

\begin{figure}[tb]
	\centering
	\includegraphics[width=0.4\textwidth]{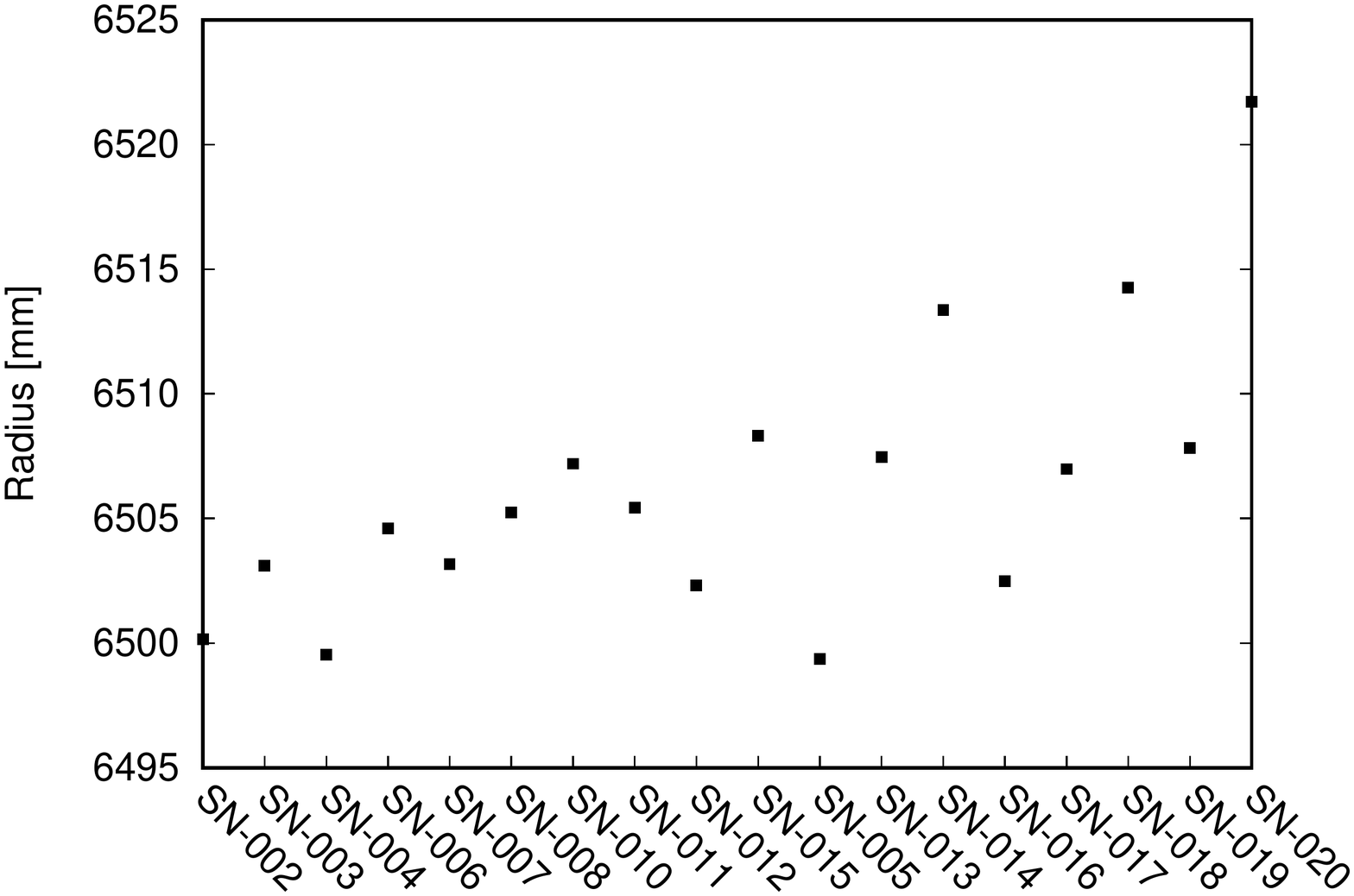}
	\includegraphics[width=0.4\textwidth]{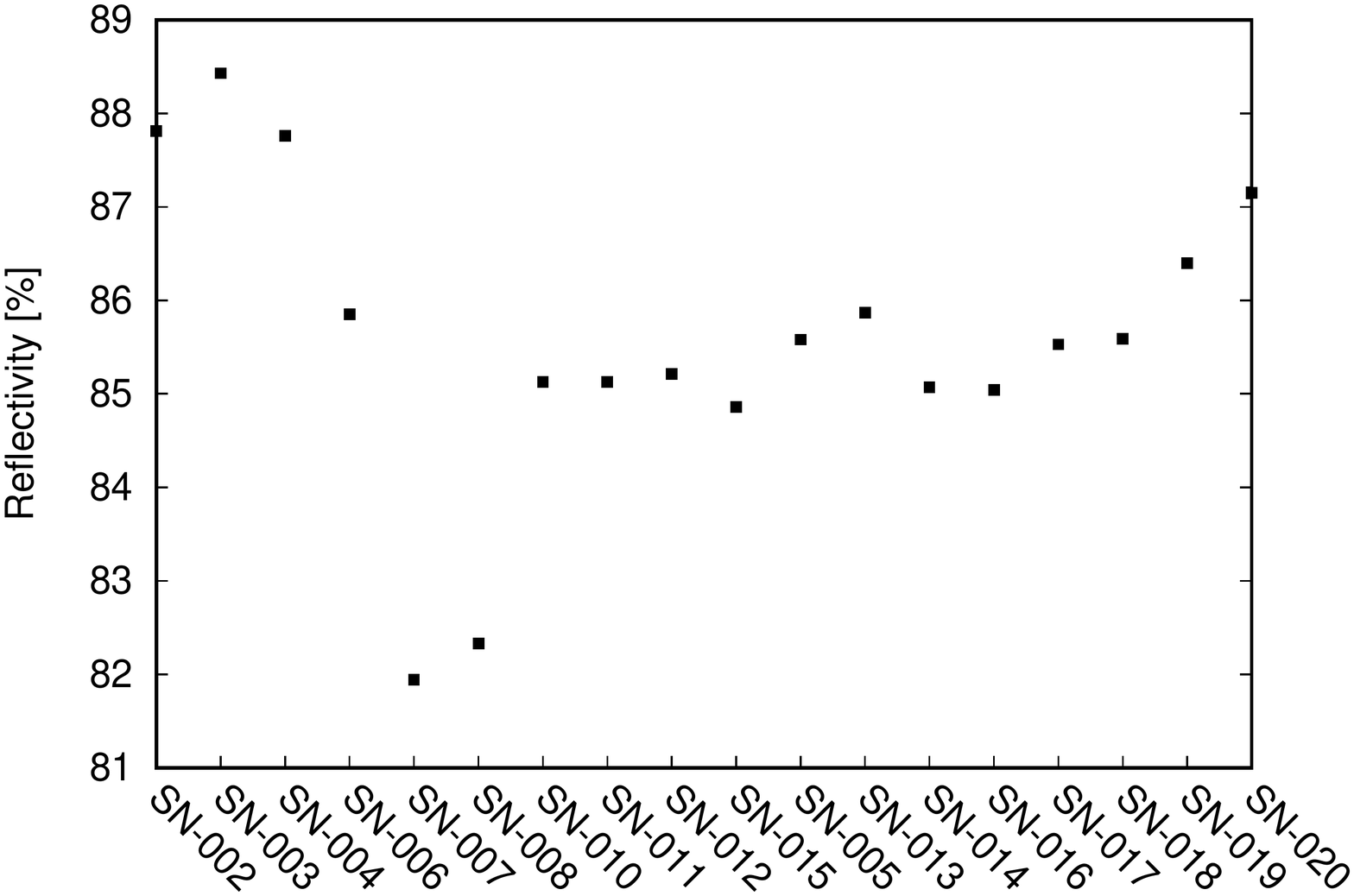}

	\includegraphics[width=0.6\textwidth]{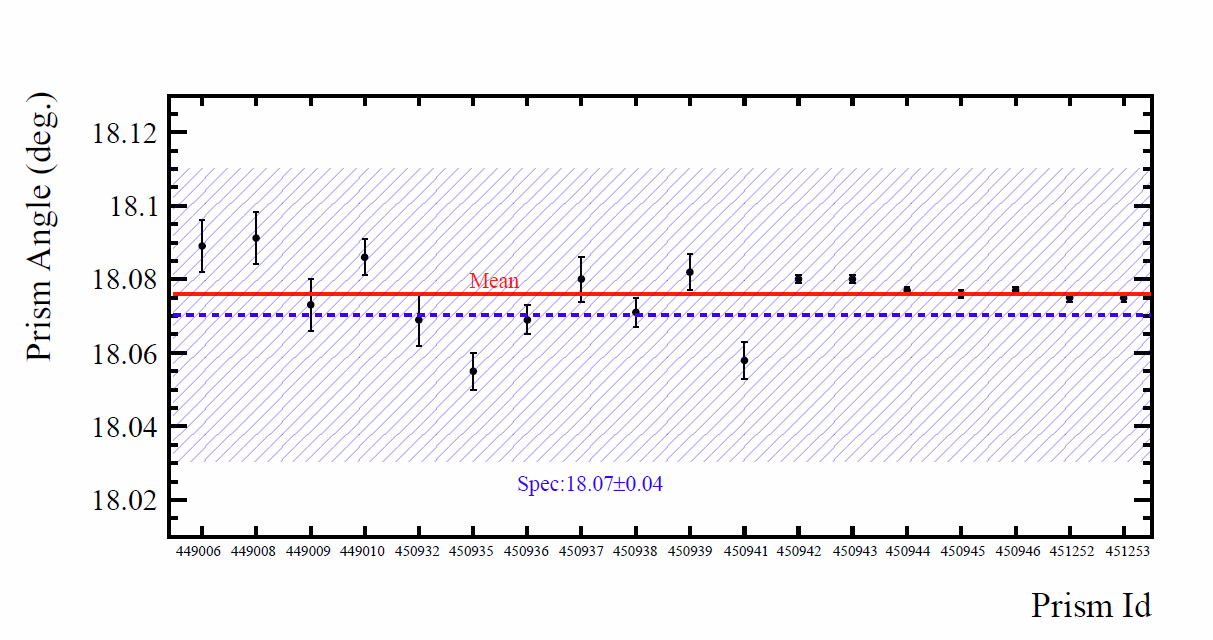}
	\caption{QA results for mirror radius (top left), mirror reflectivity (top right) and the angle of the tilted surface of the prism (bottom).}
	\label{fig:prism-mirror-qa}
\end{figure}

The mirrors and prisms were delivered to University of Cincinnati in
US after procurement. After the QA testing, they were delivered to KEK
for module assembly. The angle of the tilted surface of the prism and
the radius and reflectivity of the mirror's spherical surface were the
most important for the QA testing. They were measured by injecting
laser beam to the optics and measure the laser direction after it went
through or reflected by the optics. The results are summarized in
Figure~\ref{fig:prism-mirror-qa}. All optics fulfilled the
requirements.

\subsection{Alignment and Gluing}

After the QA process was finished and all the quartz optics passed the
requirements, two quartz bars, one prism and one mirror were mounted
on a gluing stage for precision alignment and gluing.

\begin{figure}[tb]
	\centering
	\includegraphics[width=0.6\textwidth]{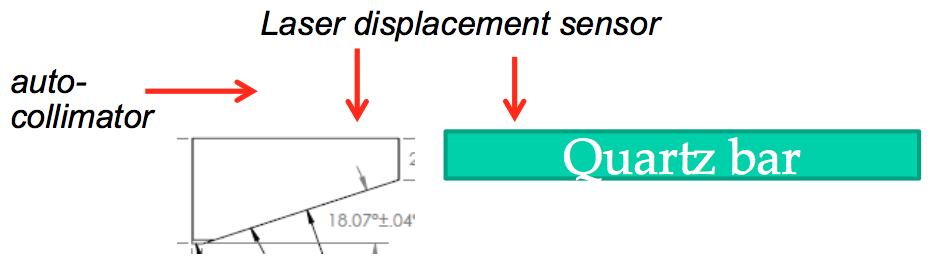}
	\caption{The alignment of the optics.}
	\label{fig:align}
\end{figure}

Two laser displacement sensors and an autocollimator were used for the
alignment as shown in Figure~\ref{fig:align}. Laser displacement
sensor was used to align the horizontal and vertical positions of two
optics' surfaces. Autocollimator accompanied with a mirror mounted on
the optics was used to align the relative angle, both horizontal and
vertical between two optics' surfaces. 


After the alignment, the optics were moved closer with a 50 $\sim$ 100
$\mu m$ gap. The joints between two optics were taped on 3 of 4 sides
by using Teflon tape to make a ``dam'' to prevent the epoxy from
flowing outside. After the mixture and centrifuge of the two parts of
the epoxy, EPOTEK 301-2, the adhesive was applied from a syringe to
the glue joint by using high pressure dry air. After applying, it took
3 $\sim$ 4 days to be fully cured.

After curing, the excessive glue was removed by using Acetone. The
alignment may have changed during the curing process, so it needed to
be measured again. The achieved horizontal and vertical angle between
the two optics near the glue joint was within $\pm$ 40 arcsec and
$\pm$ 20 arcsec, respectively.

\subsection{Module Assembly}

\begin{figure}[tb]
	\centering
	\includegraphics[width=0.6\textwidth]{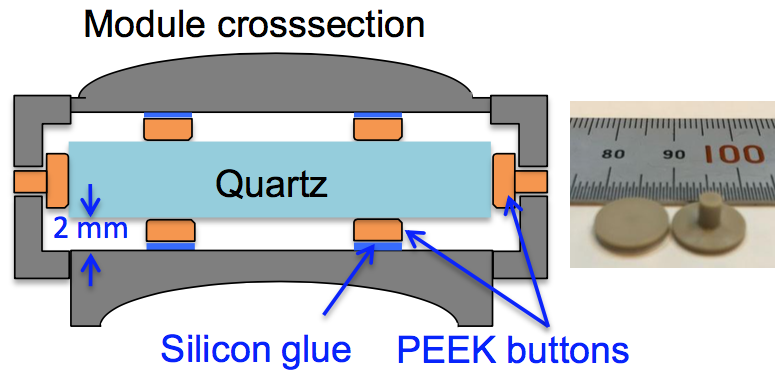}
	\caption{The cross section of the QBB (left) and the PEEK button (right).}
	\label{fig:qbb}
\end{figure}

The completed iTOP optics was moved into a Quartz Bar Box (QBB). The
QBB consisted of honeycomb panels held together with thin aluminum
side-rails. On the inner surfaces were attached PEEK buttons, which
supported the quartz optics (see Figure~\ref{fig:qbb}). The height of
the PEEK buttons were tuned precisely according to the alignment of
the optics. The QBB was attached to a support structure called a
"strong back", with which the module sag can be kept below 0.5 mm.

\begin{figure}[tb]
	\centering
	\includegraphics[width=0.4\textwidth]{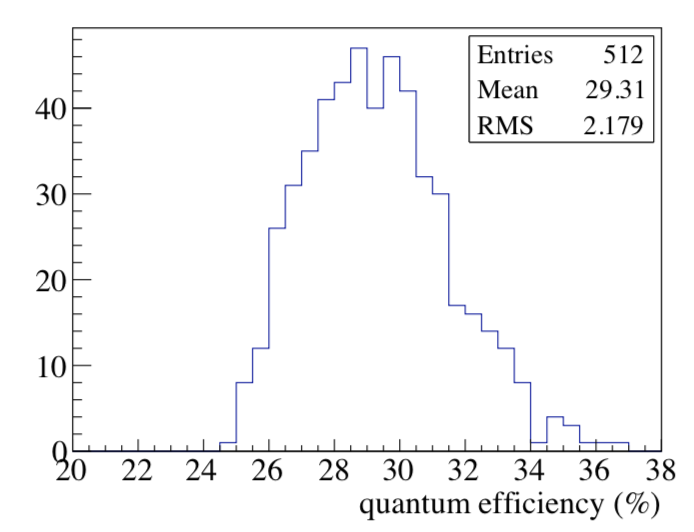}
	\includegraphics[width=0.4\textwidth]{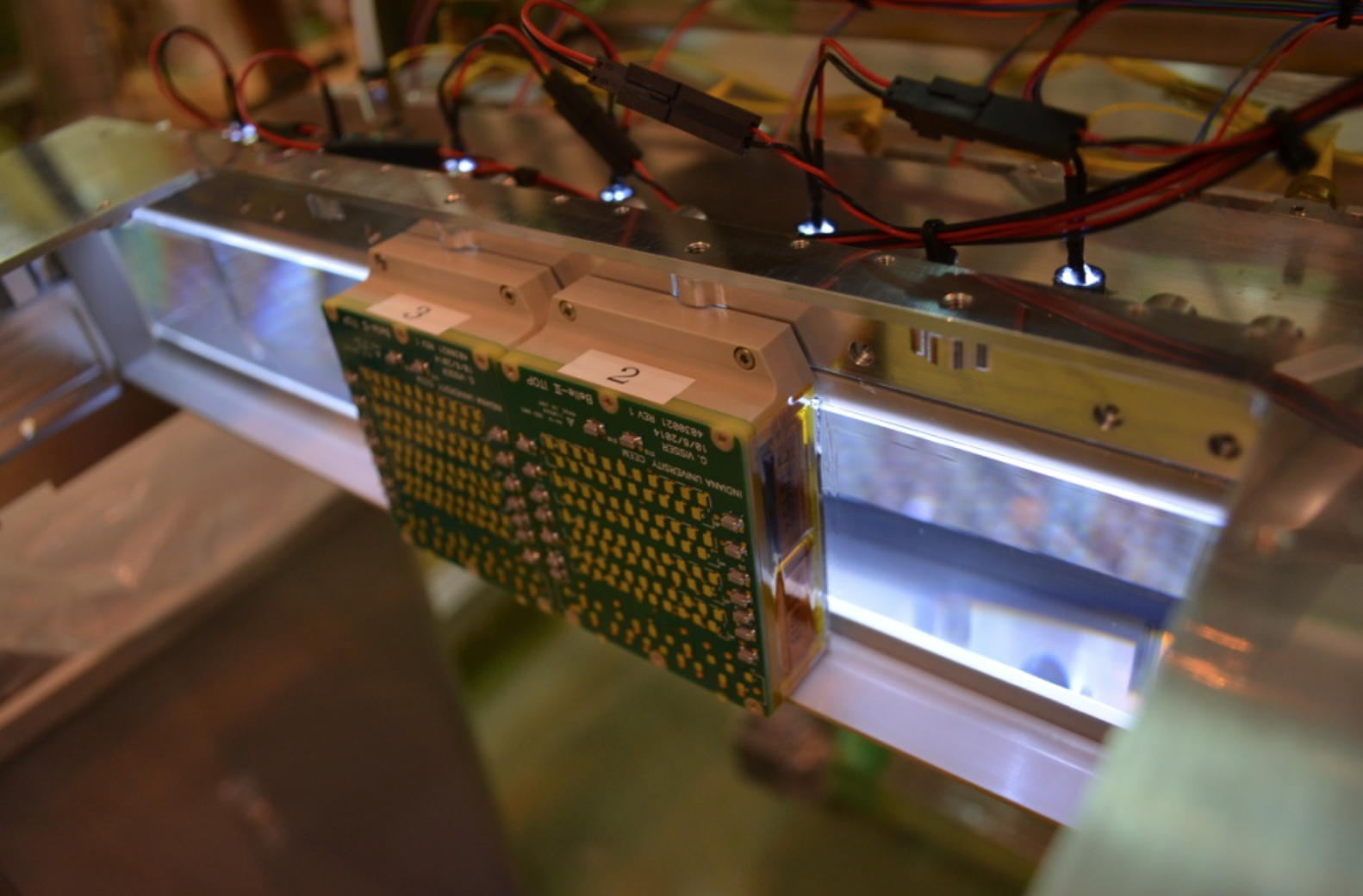}
	\caption{The quantum efficiencies for all MCP-PMTs (left) and the installation of the PMT modules (right).}
	\label{fig:pmt}
\end{figure}

The micro-channel-plate photo-multiplier tubes (MCP-PMT) Hamamatsu
SL10 is the photon sensor for the detection of the Cherenkov photons.
It has enough gain to detect single photons in 1.5T of the Belle~II
magnet field. The quantum efficiencies of all MCP-PMTs installed to
iTOP modules are shown in Figure~\ref{fig:pmt}. The average quantum
efficiency is 29.3\%, which is much better than the requirement of
24\%.

After the completion of the QBB assembly, the PMT modules, which
consist of four MCP-PMTs for each, were installed on the prism surface
of the module, as shown in Figure~\ref{fig:pmt}.

\begin{figure}[tb]
	\centering
	\includegraphics[width=0.6\textwidth]{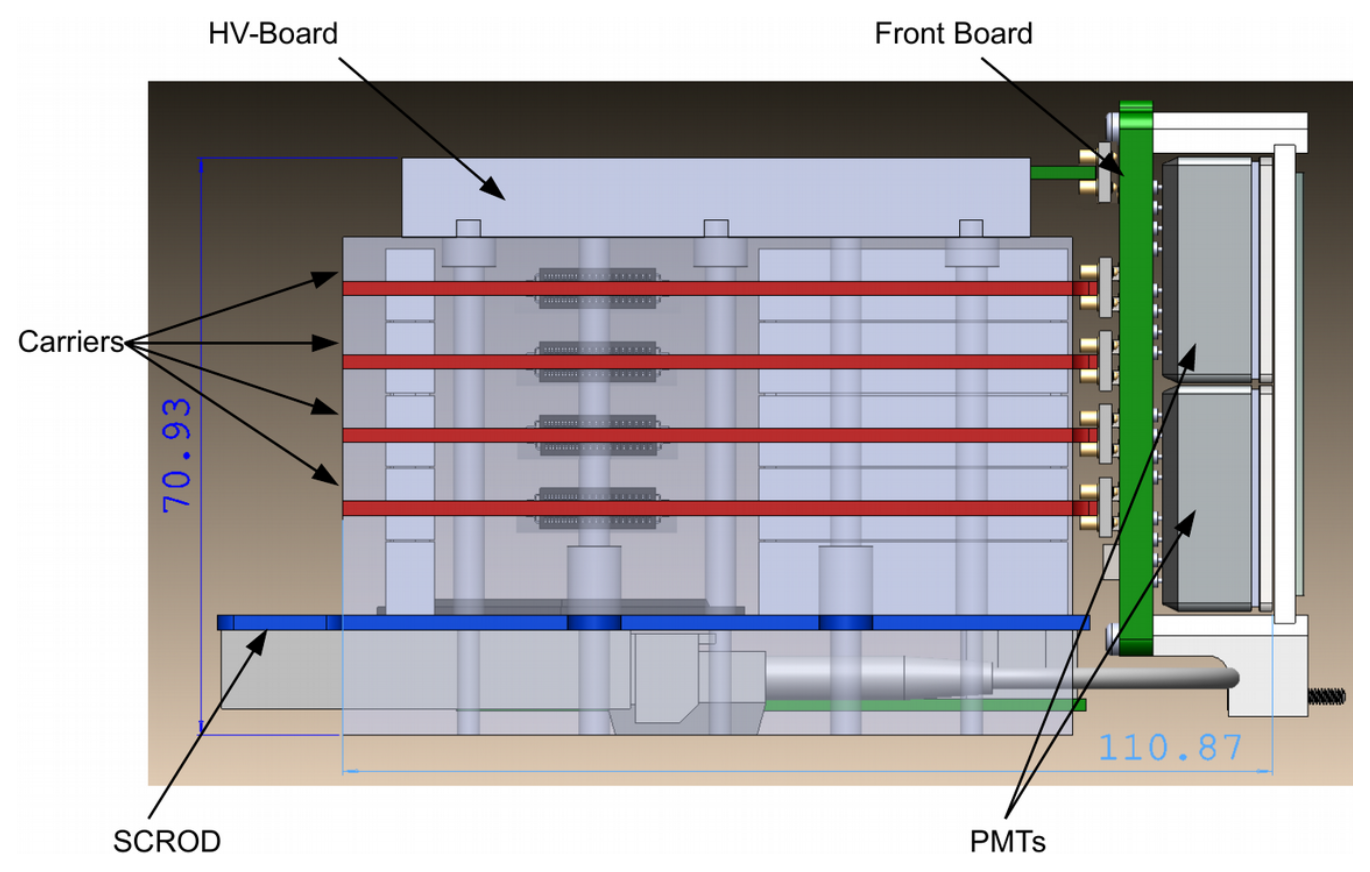}
	\caption{The structure of the board-stacks.}
	\label{fig:boardstack}
\end{figure}

The front-end read-out electronics consists of eight-channel, multi-
giga sample per second, transient waveform sampler ASICs and the FPGAs
that control the ASICs and process the data. Together with the HV
boards, they constitute the board-stacks that were installed in the
module as shown in Figure~\ref{fig:boardstack}.

\subsection{Installation}

\begin{figure}[tb]
	\centering
	\includegraphics[width=0.4\textwidth]{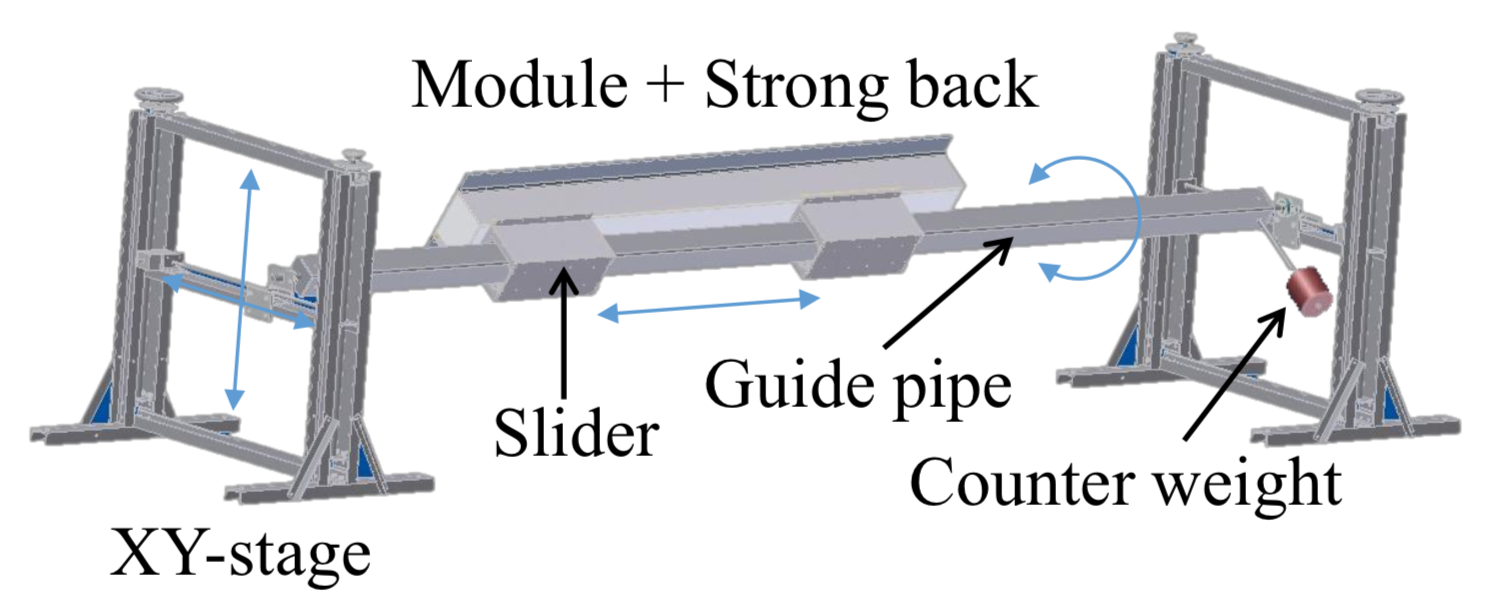}
	\includegraphics[width=0.4\textwidth]{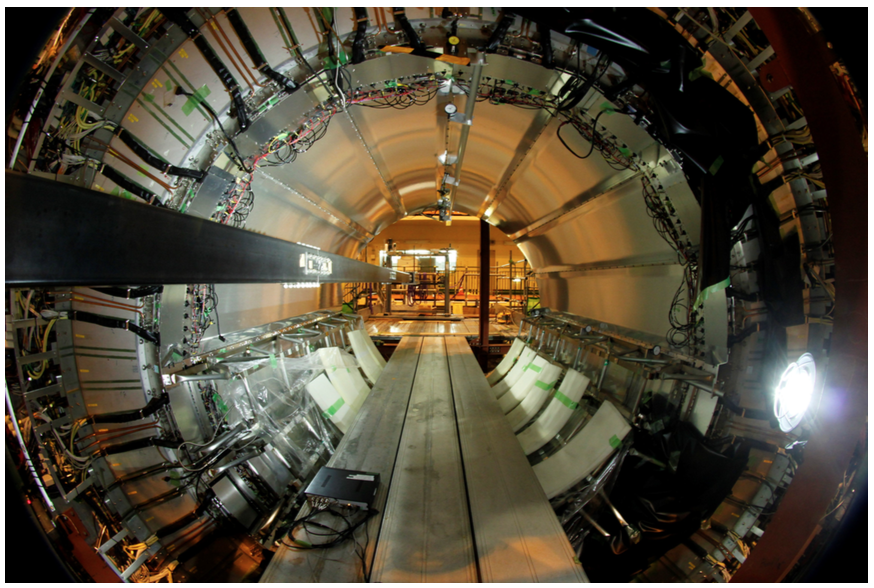}
	\caption{The installation stage (left) and the installed modules (right).}
	\label{fig:install}
\end{figure}

After the construction and testing with cosmic ray and laser,
the iTOP modules were transferred by truck to the experimental hall
for installation. A specially designed movable stage was used for the
installation, as shown in Figure~\ref{fig:install}. The module to be
installed was mounted on a guide pipe, which was supported by the x-y
stages on the two ends. The module was able to move and rotate along
the guide pipe. The module deflection during the installation process
was monitored by deflection sensors, and it was required to be less
than 0.5 mm. The installation of all the modules was completed in May,
2016. The 16 installed modules are shown in Figure~\ref{fig:install}.
More details can be found in Ref.~\cite {itop-install}.

\section{Commissioning with Cosmic Ray}

\begin{figure}[tb]
	\centering
	\includegraphics[width=0.6\textwidth]{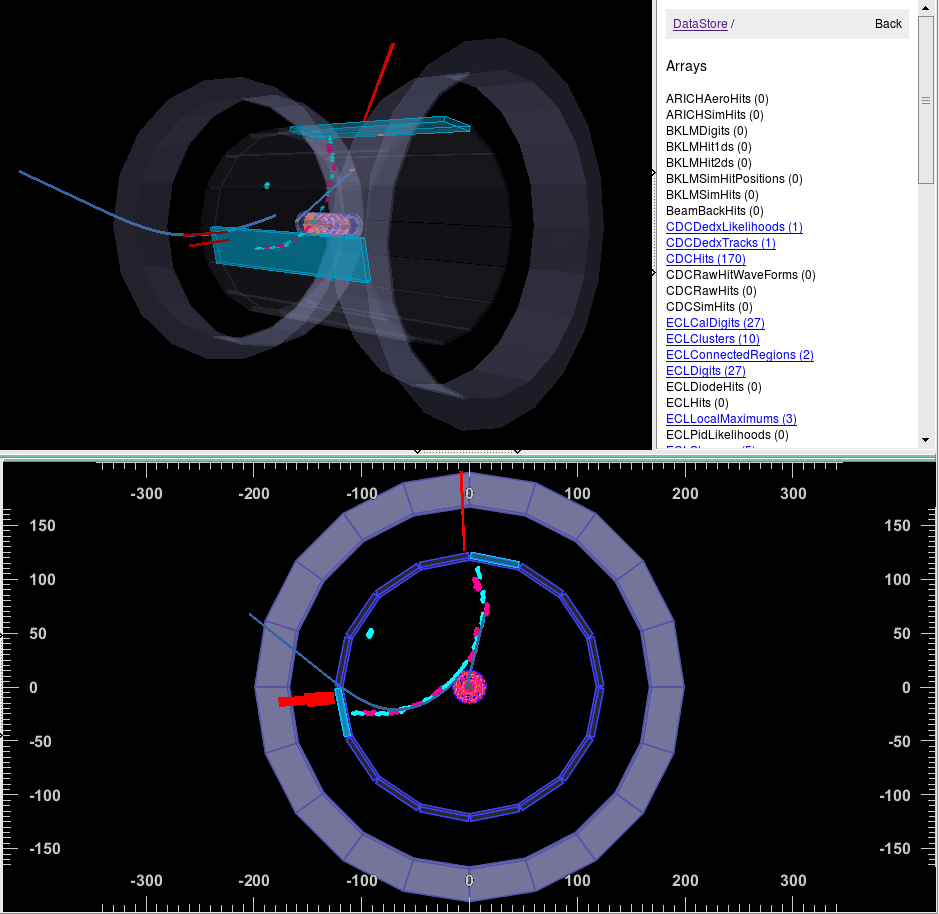}
	\caption{One cosmic ray track in the event display.}
	\label{fig:daq}
\end{figure}

After the installation of the iTOP modules, laser data and cosmic ray
data have been taken in order to debug the firmware and the local iTOP
DAQ (data acquisition) system. When other detectors such as CDC
(Central Drift Chamber) has been installed, a joint cosmic ray data
taking with all installed sub-detectors, including CDC, TOP, ECL
(electromagnetic calorimeter) and KLM ($K_L$ and muon detector),
started from the beginning of July, 2017 and lasted for around 2
months. During this period, various components of the global DAQ
system has been tested and confirmed working. An event display of one
cosmic ray track is shown in Figure~\ref{fig:daq}. From this figure a
cosmic ray track clearly went across two iTOP modules, while it was
bending inside the magnetic field and leaving hits in the CDC and ECL.

\section{Summary}

The iTOP counter is an important particle identification device for
the Belle~II detector. Here we described the design, construction and
commissioning of the iTOP counter. The last iTOP module was completed
and installed in May 2016, and the Belle~II detector was moved into
the beam line in April 2017. The global cosmic ray data taking started
in the beginning of July and lasted for around two months. Now other
sub-detectors, such as ARICH (Aerogel Ring-Imaging Cherenkov detector)
and vertex detectors, will be installed soon and the physics data
taking will be started around late 2018.

\def\Discussion{
\setlength{\parskip}{0.3cm}\setlength{\parindent}{0.0cm}
     \bigskip\bigskip      {\Large {\bf Discussion}} \bigskip}
\def\speaker#1{{\bf #1:}\ }
\def\endDiscussion{}

\end{document}